%
\input darmacros.txm	
\input epsf.tex		


\centerline{\bf Pinholes may Mimic Tunneling}
\bigskip
\centerline{D.A.\ Rabson}
\centerline{\it Department of Physics, PHY 114, University of South Florida,
Tampa, FL 33620, USA}
\smallskip
\centerline{B.J.\ J\"onsson-\AA kerman,
A.H.\ Romero,$^a$
R.\ Escudero, C.\ Leighton, S.\ Kim,
Ivan K.\ Schuller}
\centerline{\it Physics Department 0319, University of California, San Diego,
La Jolla, CA 92093, USA}
\bigskip

Interest in magnetic-tunnel junctions
has prompted
a re-examination of tunneling measurements through thin insulating
films. In any study of metal-insulator-metal
trilayers, one tries to eliminate the possibility of pinholes (small areas
over which the thickness of the insulator goes to zero so that the upper
and lower metals of the trilayer make direct contact).  Recently, we have
presented experimental evidence that ferromagnet-insulator-normal trilayers
that appear from current-voltage plots to be pinhole-free may nonetheless
in some cases
harbor pinholes. Here, we
show how pinholes may arise in a simple but realistic model of film
deposition and that purely classical conduction through
pinholes may mimic one aspect of tunneling, the exponential decay in current
with insulating thickness.

\bigskip\bigskip
\bigskip
\noindent$^a$ Current address: 
Max-Planck Institut, Heisenbergstra\ss e 1, 70569 Stuttgart, Germany

\vfil
\centerline{\bf Submitted to {\bfit Journal of Applied Physics}}
\vfil
\eject

\header{Tunnel Junctions}
The construction of magnetic
tunnel junctions
with large room-temperature mag\-ne\-to\-re\-sis\-tance\prlnote{%
J.S.\ Moodera, L.R.\ Kinder, T.M.\ Wong, and R.\ Meservey, %
{\it Phys.\ Rev.\ Lett.\ \bf74\rm, 3273 (1995)}.}
has triggered an intense research effort, as groups have applied
them as magnetic-field sensors,\prlnote{%
J.M.\ Daughton, {\it J.\ Appl.\ Phys.\ \bf81\rm, 3758 (1997)}.}
memory devices,
\prlnotez{%
S.S.P.\ Parkin, K.P.\ Roche, M.G.\ Samant, P.M.\ Rice, R.B.\ Beyers, %
R.E.\ Scheuerlein, %
E.J.\ O'Sullivan, S.L.\ Brown, J.\ Bucchigano, D.W.\ Abraham, Y.\ Lu, %
M.\ Rooks, P.L.\ Trouilloud, R.A.\ Wanner, and W.J.\ Gallagher, %
{\it J.\ Appl.\ Phys.\ \bf85\rm, 5828 (1999)}.}%
\ens{Parkin}%
\prlnotez{%
S.\ Tehrani, J.M.\ Slaughter, E.\ Chen, M.\ Durlam, J.\ Shi, and
M.\ DeHerrera, %
{\it IEEE Trans.\ Magn.\ \bf35\rm, 2814 (1999)}.}%
\ens{Tehrani}%
$\!\!^{\enr{Parkin}-\enr{Tehrani}}$
and magnetic-medium read heads.\prlnote{%
J.\ Zhang, {\it Data Storage \bf5\rm, 31 (1998)}.}
The current technological drive toward a lower product of resistance and
area,
\prlnotez{%
D.\ Song, J.\ Nowak, and M.\ Covington, {\it J.\ Appl.\ Phys.\ \bf87\rm, 5197 %
(2000)}.}%
\ens{Song}%
\prlnotez{%
E.Y.\ Chen, R.\ Whig, J.M.\ Slaughter, D.\ Cronk, J.\ Goggin, G.\ Steiner, %
and S.\ Tehrani, {\it J.\ Appl.\ Phys.\ \bf87\rm, 6061 (2000)}.}%
\ens{Chen}%
$\!\!^{\enr{Song}-\enr{Chen}}$
which implies the use of thinner and thinner insulating barriers, has
re-opened the question of how to rule out the presence of pinholes,
direct metal-metal contacts through the nominally insulating barrier.
Recent high magnetoresistances of up to 300\% at room temperature observed in
magnetic nanocontacts\prlnote{N.\ Garc\'\i a, M.\ Mu\~noz, and Y.-W.\ Zhao, %
{\it Phys.\ Rev.\ Lett.\ \bf82\rm, 2923 (1999)}; %
G.\ Tatara, Y.-W.\ Zhao, M.\ Mu\~noz, and N.\ Garc\'\i a, %
{\it Phys.\ Rev.\ Lett.\ \bf83\rm, 2030 (1999)}.}
raise the intriguing question of whether conduction through pinholes
might actually contribute to the magnetoresistance of tunnel junctions.
Similarly, any study of anomalous capacitance of magnetic tunnel
junctions
\prlnotez{%
K.T.\ McCarthy, S.B.\ Arnason, and A.F.\ Hebard, %
{\it Appl.\ Phys.\ Lett.\ \bf74\rm, 302 (1999)}.}%
\ens{McCarthy}%
\prlnotez{%
S.\ Zhang, {\it Phys.\ Rev.\ Lett.\ \bf83\rm, 640 (1999)}.}%
\ens{SZhang}%
$\!\!^{\enr{McCarthy}-\enr{SZhang}}$
must take pinholes into account.
To exclude the possibility of
pinholes,
Rowell and others in the 1960s and 1970s developed a set of
criteria to distinguish tunneling from other current paths.\prlnote{%
See {\it e.g.}
E.\ Burstein and S.\ Lundqvist, eds., {\it Tunneling Phenomena in
Solids}, Plenum Press, New York, 1969.}  Four of these criteria
continue to apply when neither electrode
superconducts: (1) the exponential thickness
dependence of the current, (2) the parabolic shape of the differential
conductance as a function of voltage, (3) the scaling of the junction
resistance with area, and (4) the temperature dependence
of the conductivity.

We have recently constructed a series of junctions demonstrating experimentally
that the second criterion applied alone cannot distinguish pinhole conduction
from tunneling.%
\prlnotez{%
B.J.\ J\"onsson-\AA kerman, R.\ Escudero, C.\ Leighton, S.\ Kim, %
I.K.\ Schuller, and D.A.\ Rabson, to appear in {\it Applied Physics Letters}.}%
\ens{experimental}%
\prlnotez{%
H.\ Srikanth and A.K.\ Raychaudhuri, %
{\it Phys.\ Rev.\ B\bf46\rm, 14713 (1992)} find a change in the sign of %
curvature of differential conductance, due to local heating,
for a series of scanning-tunneling %
point contacts %
ranging from microshorts to tunnel junctions.
Their well-defined point contacts differ from our accidental pinholes.}%
\ens{Srikanth}%
$^{\enr{experimental}-\enr{Srikanth}}$
The present work will show the first criterion similarly unreliable, since
a purely classical conduction path through pinholes may under realistic
assumptions mimic the exponential thickness dependence of the conduction
current.  This leaves only the temperature and area dependences of
conductivity as
good criteria for determining whether a junction contains pinholes.

Kleinsasser {\it et al.}\prlnote{%
A.W.\ Kleinsasser, R.E.\ Miller, W.H.\ Mallison, and G.B.\ Arnold,
{\it Phys.\ Rev.\ Lett.\ \bf72\rm, 1738 (1994)}.}
have shown that another of the proposed signatures of tunneling in
superconducting-insulator-normal junctions, the subharmonic gap
structure of Andreev reflection, may similarly reflect pinhole conduction,
but the problem of flushing out pinholes from normal-metal junctions remains.

Small pinholes may be invisible to surface microscopy,
although recently ``hot spots'' have been observed using scanning tunneling
microscopy.\prlnote{%
W.\ Wulfhekel, B.\ Heinrich, M.\ Klaua, T.\ Monchesky, F.\ Zavaliche,
R.\ Urban, and J.\ Kirschner, ``Characterization of single-crystal
magnetotunnel junctions by local tunneling,'' unpublished (2000).
}
Moreover, the large resistivity ratio between an insulator and
a metal (about $10^6$--$10^8$) requires that, over the junction area, pinhole
regions of one part in $10^6$--$10^8$ must be ruled out to ensure no pinhole
conduction in parallel with tunneling.  Generally speaking, this is very
difficult to do and not commonly considered.
Recognizing the technological, as well as scientific, importance of
identifying pinholes in magnetic tunnel junctions, Schad {\it et al.}
have developed a method for imaging pinholes through decoration by
electrodeposited copper.\prlnote{%
R.\ Schad, D.\ Allen, G.\ Zangari, I.\ Zana, D.\ Yang, M.\ Tondra,
and D.\ Wang, {\it Appl.\ Phys.\ Lett.\ \bf76\rm, 607 (2000)};
D.\ Allen, R.\ Schad, G.\ Zangari, I.\ Zana, D.\ Yang, M.\ Tondra, and D.\ Wang,
{\it J.\ Appl.\ Phys.\ \bf87\rm, 5188 (2000)}.}
This tool complements criterion (4) of temperature dependence and further
highlights the risks of relying solely on thickness dependence and differential
conductance.

\header{Tunneling {\bfit versus\/} Pinhole Conduction}
In a series of
metal-insulator-metal trilayers in which the insulating thickness, $z$,
varies, the current at given voltage should decay exponentially in $z$
(the applied voltage is assumed less than the insulating gap).
Such exponential decay has been cited as experimental evidence for
good tunnel junctions.\prlnote{%
R.\ Meservey, P.M.\ Tedrow, and J.S.\ Brooks, {\it J.\ Appl.\ Phys.\ \bf53},
1563 (1982).}
By the WKB
approximation, the characteristic decay length equals
$$
z_0 = {\hbar\over{2(2 m \phi)^{1/2}}}\smash{\quad,}
\eq(z0)
$$
where $m$ is an effective mass and $\phi$ a potential energy on the order of
the barrier height.  For $\phi=1/4$ eV and $m$ the bare electron mass,
$z_0 \sim 2$\AA.  Notably, for realistic parameters $\phi$ and $m$, this
decay length coincides roughly with the
thickness of a single atomic layer.
In contrast, a classical
resistor supports a current inversely proportional to the thickness.

The inverse relationship holds for a perfectly even layer, but
real deposition processes leave an uneven insulator with possible pinholes.
The simplest model for classical conduction by pinholes gives an exponential
dependence of resistance on deposited thickness, mimicking quantum
tunneling.  Consider a metallic substrate on top of
which we randomly deposit cubes of a perfect insulator to
an average height $\mu$ (measured in monolayers).
This is not a uniform height, so there may be pinholes.  We then deposit
a perfect conductor, making contact through any pinholes with the metal
substrate.  Since the insulator and overlayer are perfect, conduction is
directly proportional to contact area and inversely to the metal-metal
contact resistance, $R_0$.

We consider deposition to take place on a regular
two-dimensional grid, each cell of
which may be occupied by
any non-negative integral height
($0$, $1$, $\dots$)
of
insulating particles.  If each deposited particle can land
randomly over any grid cell, the resulting heights follow a Poisson
distribution.  In the
limit of an infinite two-dimensional grid, the probability that any
given grid cell contains no insulating particles is $\exp(-\mu)$.  For
a large system, this is also the expected proportion of cells that
will be unoccupied and so proportional to the conduction.
Thus where tunneling can
lead to an exponential-decay length of a monolayer for
certain realistic parameters, the simple classical model {\it always\/}
gives a decay length of one layer.

This trivial model (perfect insulating blocks) predicts conductance
that decays exponentially in coverage $\mu$ for all coverages.  However, one
expects
a crossover to $1/\mu$ decay at larger coverages.
The simplest extension of the model that might display such behavior
deposits blocks randomly, but the insulators now have some finite
resistance, $R$.  In the regime of interest, $R\gg R_0$.  As before,
we deposit $M$ blocks over an $L\times L$ lattice for a coverage $\mu=M/L^2$
(with length measured in units of monolayer thickness).
A face between the metal
overlayer and an insulating block has resistance $R$, as does a face between
two insulating blocks.
Any connection to ground has contact resistance $R_0$.
To make a simple model, we first
turn off sideways conduction in the insulators,
reducing the problem to that of an ensemble of independent
columns of binomially-distributed resistors in parallel.  We compare
this ``independent-column'' model to numerical calculations below.
Finally, we shall consider numerically an isotropic model that realistically
models growth. 

\def\ave#1{\!<\!\!#1\!\!>}		
\def\bigave#1{\left<#1\right>}		
\def\eqpunc#1{\smash{\quad#1}}		
\def\oneFonefunc{{{}_1\!F_1}}
\def\oneFone#1#2#3{{\oneFonefunc(#1,#2;#3)}}	

\header{Independent-column model}
Each independent column contains $n$ resistors of value $R$, where $n$
is the column height, and one contact resistor of value $R_0$.  Letting
$s=R_0/R$,
we have for the average column conductance
$
\ave{\sigma} = \bigave{1\over{(n+s)R}}.
$
In the Poisson limit, $L^2\rightarrow\infty$, the scaled conductance
$$
\eqalign{
S = R\ave{\sigma} &= e^{-\mu} \sum_{n=0}^{\infty} {{\mu^n}\over{n!(n+s)}}\cr
&= e^{-\mu} \mu^{-s} \int_0^\mu t^{s-1} e^t dt
~=~ {1\over s} e^{-\mu} \,\oneFone{s}{s+1}{\mu}\eqpunc.\cr
}
\eq(Rsigma)
$$
Here, $\oneFonefunc$ is a confluent hypergeometric function.

One arrives at an asymptotic series for large $\mu$ through
successive integrations by parts,\prlnote{%
C.M.\ Bender and S.A.\ Orszag, {\it Advanced Mathematical
Methods for Scientists and Engineers}, McGraw-Hill, New York, 1978.}
exhibiting
explicitly the approach to $1/\mu$ conductance:
$$
R\ave{\sigma} ~\sim~ \mu^{-1} \left[ 1 + \Gamma(s) \sum_{j=1}^n
{{(-1)^j}\over{\Gamma(s-j)\mu^j}} \right]\eqpunc,
\eq(Rsigma:asymp)
$$
where $n$ is a cutoff that must be introduced for any finite $\mu$.

More interesting is the small-$\mu$ limit, in which we recover the
exponential decay of the trivial model, for multiplying the top equation in
\(Rsigma) by $s$, we have
$$
R_0\ave{\sigma} = e^{-\mu}\left(1 + s\sum_{n=1}^\infty {{\mu^n}\over{n!(n+s)}}
\right)
\eqpunc.
\eq(R0sigma)
$$
For small enough $\mu$,
$R_0\ave{\sigma} \sim e^{-\mu}$, representing exponential decay of
conductance with coverage.

We can estimate the crossover scale $\mu_0$ above which the decay
ceases to be exponential by setting the ``1'' term in parentheses in
\(R0sigma) equal to the remainder.  This yields the condition
$$
2 = \oneFone{s}{s+1}{\mu_0}\eqpunc.
\eq(crossover1)
$$
For $s\!\leo\!0.5$, we can
replace $\oneFonefunc$ with the leading term in its 
asymptotic expansion,
giving
$$
\mu_0 \approx \ln(2/s)\eqpunc.
\eq(crossover2)
$$
The logarithmic form of \(crossover2) comes as something
of a surprise: it means that the non-zero contact resistance cuts off the
small-coverage, exponential regime no matter how tiny $R_0$ is relative to
the insulating $R$.

\midinsert 
\centerline{\epsfxsize=0.6\hsize\epsfbox{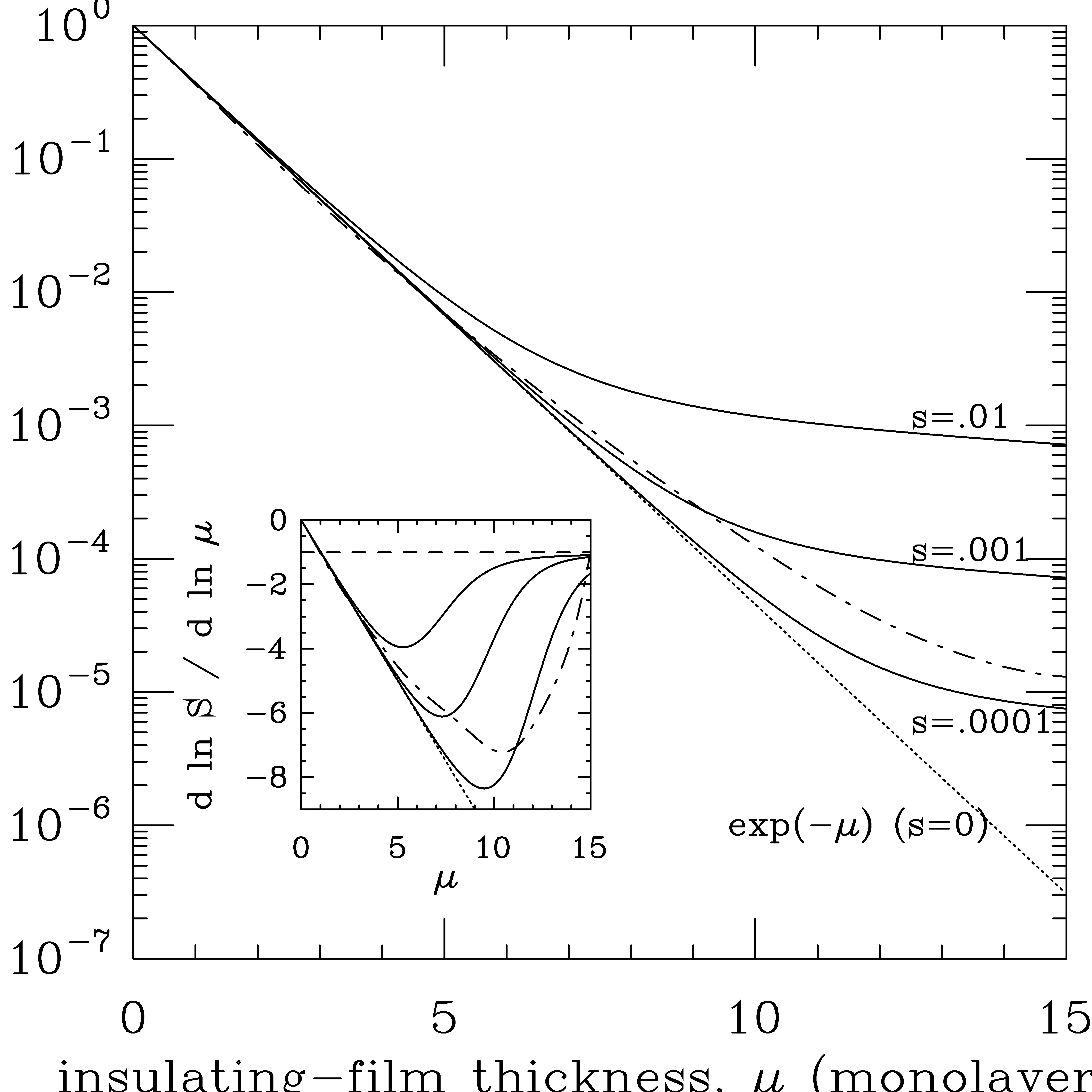}}	
\caption{
F{\ninepoint IGURE} 1.
Scaled conductances (solid traces)
given by \(Rsigma)
for three values of the ratio $s$ of contact to insulating resistance.
According to the estimate \(crossover2), they begin to cross over from
exponential to inverse-thickness behavior at about 5.3, 7.6, and 9.9.
For comparison, we define the heuristic measure $\mu_{15}$ as the point at
which the slope deviates from $-1$ by 15\%; the corresponding estimates are
4.4, 7.2, and 9.8.
The dotted line ($s=0$) is an exact exponential.  The dot-dashed line
shows a numerical simulation ($100\,000$ trials
on $5\times5$ grids) for $s=0.0001$
with
$J=-0.5$, diffusion $x=1/3$, and ``up''-steps allowed.
The inset plots the same curves in a way that makes explicit the approach
to $1/\mu$ conductance ($d\ln S/d\ln\mu=-1$, horizontal line) as well as the
initial exponential regime.  Note that surface relaxation in this case causes
a more rapid deviation from exponential decay ($\mu_{15}=6.1$)
while also delaying the
onset of $1/\mu$ conductance.
}
\endinsert 

The solid traces (for different $s$) in the semilog plot,
figure 1, clearly show
the exponential
regime for small $\mu$ and the gradual deviation
as pinhole conduction ceases to dominate.
To display large- and small-$\mu$ behaviors together,
it is helpful to plot $y = d \ln S/d \ln \mu = S' \mu / S$ against $\mu$;
the exponential regime is characterized by a constant slope, while in
the asymptotic $1/\mu$ regime, the graph approaches the constant $y=-1$
(inset to figure 1).
In this independent-column model, the minimum
of the graph corresponds closely to the estimate \(crossover2);
later, in numerical simulations incorporating surface relaxation, we see 
deviations somewhat earlier.

We have assumed the ratio $s$ small,
as supported by the following estimate.
Using Sharvin's semiclassical calculation\prlnote{
Yu.\ V.\ Sharvin, {\it Zh.\ Eksp.\ Teor.\ Fiz.\ \bf48}, 984 (1965)
[{\it Sov.\ Phys.\ JETP \bf21}, 655 (1965)].}
for contact
resistance through a pinhole of diameter $\sim1$\AA, and assuming
a typical metallic Fermi temperature and electronic density, we
have contact resistance $R_0\sim10^4\Omega$.
The ``classical'' resistance of an insulating block is less well defined,
but a minimum resistivity $\sim10^6\Omega$ cm suggests
$R\!\geo\!10^{14}\Omega$, so that
$s=R_0/R\leo10^{-10}$, tiny indeed.  We then estimate by \(crossover2)
that insulating thicknesses up to about 24 monolayers should show exponential
resistance.

\header{Numerical Simulations}
The preceding model 
provides a baseline for our simulations.
In addition to restoring conduction through the
sides of blocks, we adopt a model due to Pal and Landau
\prlnotez{S.\ Pal and D.P.\ Landau, {\it Phys.\ Rev.\ B\bf49}, 10597 (1994).}%
\ens{Pal94}%
\prlnotez{D.P.\ Landau and S.\ Pal, {\it Thin Solid Films \bf272}, 184 (1996).}%
\ens{Landau94}%
\prlnotez{D.P.\ Landau and S.\ Pal, {\it Langmuir \bf12}, 29 (1996).}%
\ens{Landau96}%
$\!\!^{\enr{Pal94}-\enr{Landau96}}$
to describe the motion of insulating blocks along the surface after they have
fallen.  Such models have found
good agreement with experiment for
long-wavelength and long-time (large-$\mu$) features,
although our interest lies in the opposite regime.

For simplicity, we imagine the original metallic layer to be flat.\prlnote{%
One might also model the effect of polishing scratches in
seeding pinholes by imposing a surface potential
on the substrate; see D.J.\ Keavney, E.E.\ Fullerton, and S.D.\ Bader,
{\it J.\ Appl.\ Phys.\ \bf81}, 795 (1997).}
After depositing some number
of blocks in the manner of the previous section, we allow a fraction of
blocks
at the interface to diffuse one unit along the surface.
We employ a quasi-Metropolis procedure to determine whether to accept or
reject each move (the system is not at equilibrium) with an
energy $J$ for each face-to-face bond between blocks: calling $D$ the change
in the number of bonds, the move is accepted unconditionally if it lowers the
energy and otherwise with probability $\exp(-D J/k_{\hbox{\eightpoint B}} T)$.
Henceforth, we scale $J$ by setting the thermal energy
$k_{\hbox{\eightpoint B}} T=1$.  Thus
a negative $J$ encourages clustering, while a positive $J$ could represent
stearic hindrance or an affinity for roughness.  We adopt one more feature
from the Pal-Landau models: blocks either may or may not jump ``up'' (further
away from the substrate) while relaxing.$^{\enr{Landau96}}$

We perform the simulation on an $L\times L$ grid, measuring conductance at
set times, before averaging over
many trials to reduce noise.
Since long-wavelength features do {\it not\/}
concern us in the present work, $L$ can be as small as allowed by diffusion.
Consider the diffusion parameter $x$ defined as the fraction of interface
blocks diffusing one lattice spacing per monolayer deposited.
A simple random-walk argument establishes that for $x$ of order unity,
we can look at coverages as great as about 25
for a $5\times5$ grid.  Numerically, we tested values of $L$
from 5 to 60 and found no significant differences in conductances.
The actual diffusion parameter $x$ of course will depend on temperature and
physisorption and chemisorption energies.
We considered diffusion parameters $x=$0.33--10, roughly corresponding to
a fast, but not unrealistic, deposition rate.\prlnote{%
D.L.\ Smith, {\it Thin-Film Deposition: Principles and Practice},
McGraw-Hill, New York, 1995.}

This model must behave the same as the independent columns in the two
limits of zero and large coverage, but
as figure 1 shows, surface relaxation can reduce the thickness
at which deviations from exponential decay first become apparent, primarily by
modifying the rate at which pinholes fill in.
Because the deviation
occurs continuously, we need a heuristic measure.  The slope of the
line in a semilog plot (figure 1) determines the decay length; in both
the simple model and the simulations, it always starts out at $-1$.
Such a plot appears to differ by eye from a straight line
when the slope changes by 15\%; we call this crossover scale $\mu_{15}$.
As noted in the caption to figure 1, it also agrees reasonably well with the
analytical crossover estimate \(crossover2) for the independent-column model
with no relaxation.

\midinsert 
\centerline{\epsfxsize=0.6\hsize\epsfbox{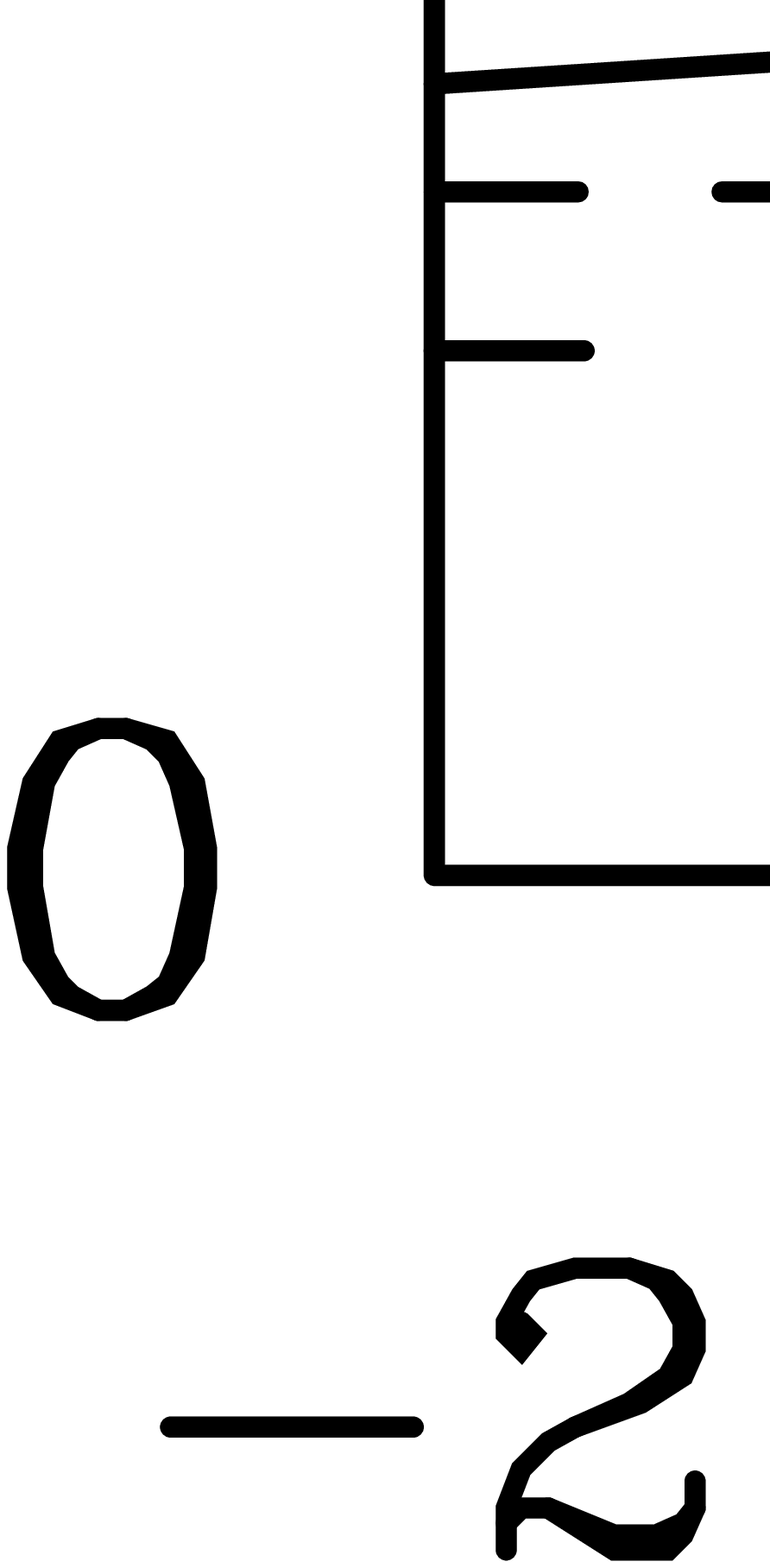}}	
\caption{
F{\ninepoint IGURE} 2.
The horizontal axis measures the bond strength (negative attracts blocks),
the vertical an estimated maximum thickness for the mimicry of tunneling,
specifically
the empirical measure $\mu_{15}$ at which the slope of a semilog
plot, as in figure 1, deviates by $15\%$ from $-1$.
For comparison,
the cutoff estimates \(crossover2) without relaxation
for $s=.01$, $.001$, and $.0001$ are
5.3, 7.6, and 9.9.  Blocks may jump up in these traces.
}
\endinsert 

Figure 2 plots $\mu_{15}$ as a function of bond strength, $J$,
for a few selected values of $s$ and the relaxation parameter, $x$.
We find that relaxation still permits a large value for $\mu_{15}$ for
particular ranges of $J$.
When blocks are permitted to jump up, there is a tendency for pinholes to
be extinguished less quickly than exponentially (see dot-dashed curve in
figure 1), since at moderate coverage
more columns of height 1 turn into pinholes than {\it vice-versa}.
A sufficiently attractive (negative) $J$ counters this tendency by
discouraging jumps out of height-1 columns, but too large a negative
value of $J$ leads to superexponential decay.  In the model without jumps,
a sufficiently large positive (repulsive) $J$ effectively cancels the effect
of surface relaxation on pinholes.

\topinsert 
\centerline{\epsfxsize=0.6\hsize\epsfbox{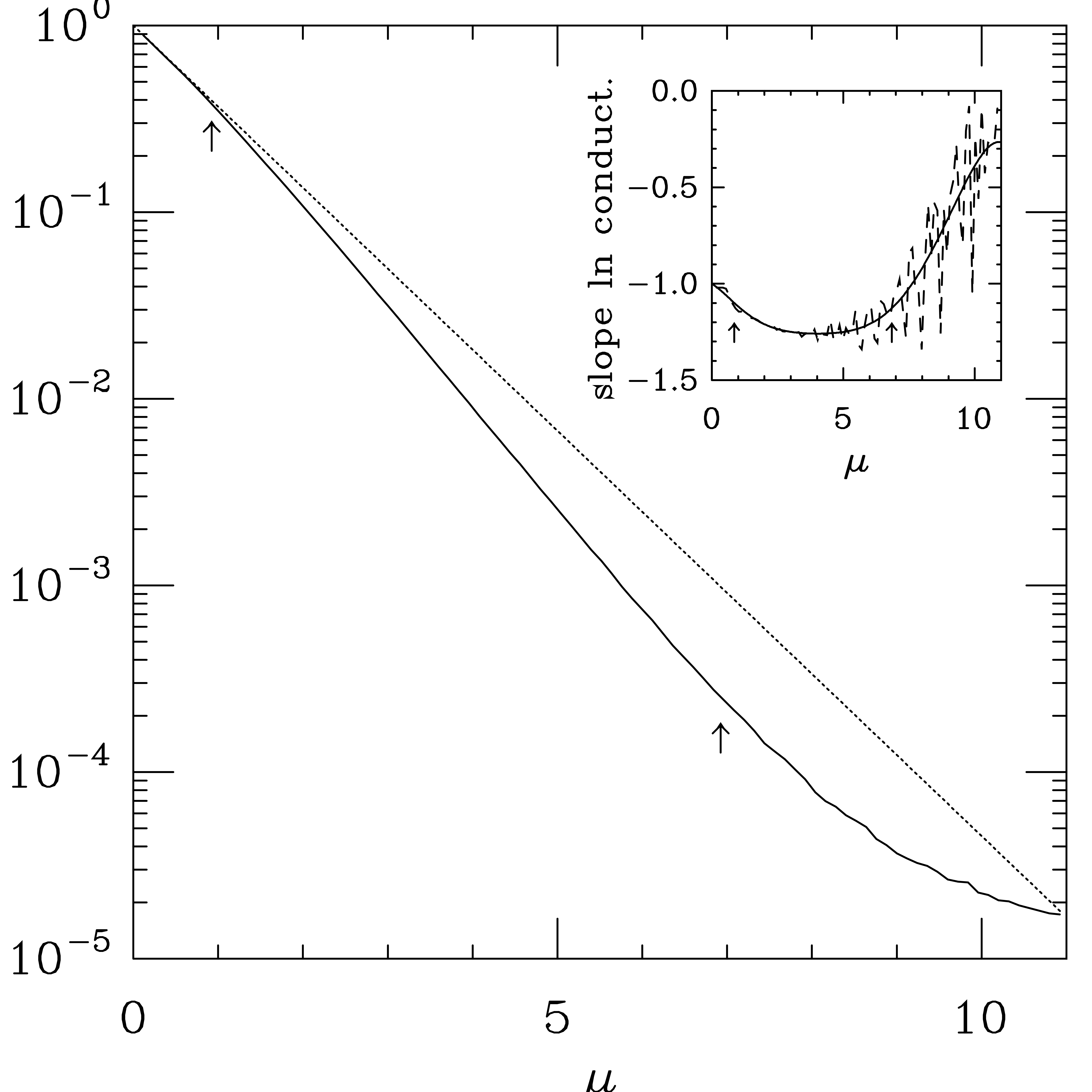}}	
\caption{
F{\ninepoint IGURE} 3.
An example of a different effective slope: the dotted line is an exact
exponential.  Between the arrows
($\uparrow...\uparrow$ $\mu\approx1...7$), it is
difficult to distinguish the simulated decay
with relaxation ($s=.0001$, $x=1/3$,
$J=-2$) from a straight line of slope $\sim -1.25$, {\it i.e.}, a decay
length of 0.8 monolayers.  No smoothing has been applied to the main graph.
The inset shows the slope of the decay; the solid line applies parametric
smoothing.
}
\endinsert 

In some cases, we find that a reasonably large negative $J$
can effectively alter the exponential-decay length over some
range of coverages; see figure 3.  To get a decay length substantially
different from one monolayer, we need to modify the
model.  For example by prohibiting direct deposition on top of
isolated pinholes (as by stearic hindrance), filling these only by
surface relaxation, we compute both a longer decay length and a
larger crossover scale (not shown).

\header{Implications for experiment}
We have concentrated on one signature of tunneling, 
showing that purely classical pinhole conduction can mimic the
exponential dependence of resistance on barrier thickness.
We have recently demonstrated
experimentally that another signature, the nonlinear form of current
$I(V)$ (as a function of bias $V$)%
\prlnotez{%
J.G.\ Simmons, {\it J.\ Appl.\ Phys.\ \bf34}, 1793 (1963).}%
\ens{Simmons}%
\prlnotez{%
W.F.\ Brinkman, R.C.\ Dynes, and J.M.\ Rowell, {\it J.\ Appl.\ Phys.\ \bf41},
1915 (1970).}%
\ens{BDR}%
$^{\enr{Simmons}-\enr{BDR}}$
also may fail to distinguish
classical conduction from tunneling.$^{\enr{experimental}}$
In a series of
ferromagnet-insulator-metal junctions in which the ``metal'' is actually
a superconductor, all samples could be fit well to a tunneling form
above the superconductor's transition temperature, $T_c$.
However, some (but not all) showed Andreev reflection below
$T_c$, indicating the presence of pinhole shorts through the
insulator.\prlnote{%
G.E.\ Blonder, M.\ Tinkham, and T.M.\ Klapwijk, {\it Phys.\ Rev.\ B\bf25}, 4515
(1982) and G.E.\ Blonder and M.\ Tinkham, {\it Phys.\ Rev.\ B\bf27}, 112 (1983)
examine the crossover from Andreev reflection
to tunneling as the insulating barrier increases in strength.  As a practical
matter, with a typical band gap ($\sim10$ eV), the Andreev
signature should become insignificant for insulating thicknesses greater than
a few monolayers.  Thus, the presence of Andreev reflection indicates
either a pinhole or a ``near'' pinhole.
}
In other words, samples that appear (according to $dI/dV$ and thickness
dependence $I(z)$) to be good tunnel junctions may not be.

We would expect that increasing temperature
should enhance current through
a tunnel junction (effectively lowering the barrier) while suppressing
conduction through any metallic short.  In fact, we found
that the temperature dependence of current $I(T)$ at zero bias does distinguish
the pinhole-free from the shorted samples.

\bigskip

\header{Acknowledgements}
We thank H.\ Srikanth and M.\ Devoret for helpful discussions and
acknowledge the coding assistance of Mr.\ Michael Grossman.
This work was supported in part by DARPA and ONR.

\vfil\eject
\header{References}
\endnotes

\vfil\eject\end